\documentclass[twocolumn]{revtex4}

\begin{document}

\title{$|0\rangle|1\rangle+|1\rangle|0\rangle$}
\author{S. J. van Enk}
\affiliation{Bell Labs, Lucent Technologies\\
600-700 Mountain Ave, Murray Hill, NJ 07974}
\date{\today}

\begin{abstract}
I give a simple argument that demonstrates that
the state $|0\rangle|1\rangle+|1\rangle|0\rangle$, 
with $|0\rangle$ denoting
a state with 0 particles and $|1\rangle$ a 1-particle state,
is entangled in spite of recent claims to the contrary.
I also discuss new viewpoints on the old controversy about
whether the above state can be said to
display single-particle or single-photon nonlocality.
\end{abstract}

\maketitle
Every now and then I hear or read the claim that there is no 
entanglement 
in the state 
\begin{equation}\label{one}
|\psi\rangle_{A,B}=|0\rangle_A|1\rangle_B+|1\rangle_A|0\rangle_B,
\end{equation}
where $|0\rangle_{A,B}$ and $|1\rangle_{A,B}$ denote states with 
zero and one particles, respectively,
in modes $A$ and $B$.
(See, for example, Ref.~\cite{czachor};
on the other hand, see also papers \cite{other} 
that use or propose to use the same state for
teleportation, quantum cryptography or violating Bell inequalities,
or that perform tomography on a similar state.) 
The reason for that claim 
is usually
one of the following: 
\begin{enumerate}
\item One needs at least two particles 
for entanglement. 
\item 
The state of Eq.~(\ref{one}) when written in second-quantized form
as 
\begin{equation}
|\psi\rangle_{A,B}=(a_A^{\dagger}+a_B^{\dagger})|{\rm vacuum}\rangle
\end{equation}
clearly has no entanglement.
\item
The entanglement is a property
of a pathological representation of CCR/CAR algebras \cite{czachor}.
\end{enumerate}
Those reasons probably
all seem crystal clear and utterly convincing.

But here is a simple counter argument \cite{atoms}
that shows
there is in fact entanglement
in the state (\ref{one}) provided modes $A$ and $B$ are spatially 
separated \cite{modes}. Just for the argument let us
assume the particles are photons \cite{electrons}. Also
let us assume we place a cavity in each of 
the locations of the modes $A$ and $B$
and put an atom, initially in a ground state denoted by
$|g\rangle$, inside each cavity. 
There are techniques \cite{cirac}
to make sure a photon
in the proper mode will enter
the cavity and excite the atom to a particular excited state, denoted
by $|e\rangle$.
In the ideal case, this process occurs with 100\% efficiency.
Starting with the two atoms both in state $|g\rangle$
we can then generate the joint atomic state 
(where the obvious assumption is made that
the vacuum will not excite the atom)
\begin{equation}\label{ge}
|\Psi\rangle_{A,B}=|g\rangle_A|e\rangle_B+|e\rangle_A|g\rangle_B,
\end{equation} 
where $A$ and $B$ now refer to the locations of the atoms.
The joint state of the two (photonic) modes is no longer relevant or entangled
as both modes end up in the state $|0\rangle$.

In the state (\ref{ge}) there are two particles 
so that objection 1 from the above list does not apply. 
Furthermore,
no one would insist on writing
the state of the two atoms in separate cavities in a second-quantized 
form, so objection 2 would not be raised. Moreover, the atoms used 
in the above-mentioned procedure do not have to be identical at all, so
(i) Eq. (2) would not apply in any case, and (ii)
there are no problems arising from the role of quantum statistics
of identical particles in the definition of entanglement.
Finally, no one would complain about pathological representations of
any sort of algebras when discussing
(nonidentical) atoms. Thus I would say there is no doubt there is 
entanglement in the state (\ref{ge}). But since 
that state can be generated in principle, as just shown, 
from the state (\ref{one}) by {\em local} operations, 
I would conclude that the state  
$|0\rangle|1\rangle+|1\rangle|0\rangle$
must have entanglement too.
That concludes the simple argument.

Some further remarks are in order:
First, in the famously \cite{two-mode}
entangled two-mode squeezed state one has a Fock-state
expansion $|0\rangle|0\rangle+|1\rangle|1\rangle+|2\rangle|2\rangle\ldots$
in the ideal (unrealistic) case of infinite squeezing, 
but in that case no one complains about the ``vacuum term.''
Moreover, in a realistic finitely-squeezed
 two-mode squeezed state the $|0\rangle|0\rangle$
term has, in fact, 
the largest amplitude, and that part does contribute to the total
entanglement \cite{enk}. For example,
for small amounts of
squeezing the state is, approximately, $|0\rangle|0\rangle+
r|1\rangle|1\rangle$, with $r\ll 1$, which has
a small amount of entanglement
on the order of $r\log_2(r)$.
	
Second, the entanglement in the state (\ref{one}) is {\em not} between the
photon and the vacuum, but between modes $A$ and $B$. 
This point has been made in more generality (for different physical
systems, for different types of states and relative to sets of
observables) in \cite{modes}. Similarly,
in the case of the two-mode squeezed state
with small squeezing, the entanglement is {\em not} between
two photons and the vacuum: here the name of the state quite
appropriately indicates what {\em is} entangled.

Third, a different reason altogether for not attributing entanglement 
to the state (\ref{one}) under certain conditions
is given in \cite{vaccaro}.
That paper refers to the situation where the relative
phase between the two states
$|0\rangle_A|1\rangle_B$ and $|1\rangle_A|0\rangle_B$ is not well defined.
This occurs when, e.g., the two parties located at $A$ and $B$ 
do not share a reference that defines that phase (for instance, a clock or a
spatial reference frame). More precisely,
suppose Alice and Bob, to use modern parlance, share a state
\begin{equation}\label{phase}
|\psi\rangle_{A,B}=|0\rangle_A|1\rangle_B+\exp(i\phi)
|1\rangle_A|0\rangle_B,
\end{equation}
where $\phi$ is defined relative to a (possibly fictituous)
third-party reference frame; the states $|0\rangle$ and $|1\rangle$
may refer now to any types of orthogonal quantum states, be it
polarization states of single photons, states of Josephson junctions 
with different charges, or spin ``up'' and ``down'' states of electrons.
 Alice and Bob may have their
own local reference frames but the difference
between their local phases or their relative orientation
is not known to them.
Hence Alice and Bob do not know the phase $\phi$ and so
they would in fact not
assign the state (\ref{phase}) 
but rather a mixture over the unknown phase
$\phi$ to a single copy \cite{Ebits}. The description
(\ref{phase}) is used by anyone with access to the third-party
reference frame. 
In this situation Alice and Bob cannot make use of 
a single copy  \cite{Ebits}
of the state
(\ref{phase}) for teleportation or violating Bell inequalities.
In that sense, according to Alice and Bob,
there is no entanglement between Alice and Bob's 
systems $A$ and $B$ when they do not share a reference frame.
Of course, when they {\em do} share a reference frame (and in experiments
this is always explicitly or implicitly assumed), there {\em is} entanglement
(see also \cite{aha}).
Note, for example, that the above-mentioned operation involving
atoms in cavities requires a phase reference, too.
In contrast, let us note that
even in the absence of a shared
reference frame, one can still perform quantum communication
protocols and violate Bell inequalities
 by using a reference-frame invariant encoding,
as discussed in \cite{encoding}.

Fourth, in the 90s a related but different
discussion arose as
to whether nonlocality can arise from a single-particle or 
single-photon state \cite{single}. The issue then was not whether there is
entanglement in the state $|0\rangle|1\rangle+|1\rangle|0\rangle$
(apparently, there was agreement there is entanglement), but whether
an experiment using that state can demonstrate nonlocality
with just 1 particle or photon.
The idea is simply that all proposed (and in the meantime performed)
optics experiments with the state (\ref{one})
detect, at least sometimes, 
more than a single photon. In that case, it was argued,
nonlocality arises from multiparticle entanglement.
We can add some new insights to that discussion
by relating it to the role reference frames
play in quantum-communication protocols

In certain types of experiments
the shared reference
frame
is such a trivial resource that no one cares to mention it.
This applies, for instance,
to experiments using 
a spatial reference frame (the earth or
the fixed stars).
On the other hand, the role of a clock (another example of a reference frame)
in optics experiments
is inevitably, conveniently, and quite visibly, played by
lasers (e.g., Alice and Bob both having a laser, phase-locked to one
another) \cite{wiseman}. 
The confusing aspect is that in
optics experiments on Bell inequalities
photons are detected
that may originate both from the entangled state (\ref{one})
and from the phase reference laser beam. In contrast, in experiments
with a spin-entangled
electron pair or a polarization-entangled photon pair the
particles making up the spatial reference frame are {\em not} detected by 
the same detector that detects the electrons or photons.
Hence it may seem that indeed only 2 electrons or 
2 photons have to be detected.
However, this apparent distinction is not so clear:
One could argue, on the one hand, that the reference frame particles {\em are} 
detected, not by a detector but by the experimenter.
On the other hand, one could argue that in optics experiments
a different sort of clock could be used, at least in principle,
that requires no photons (say, based purely on electronics). In that case,
Bell inequalities could be violated using just the single-photon state
(\ref{one}) without more than one photodetector clicking.

In fact, this is a good example of the difference
between ``internal'' and ``external'' reference 
frames \cite{refframes}: In optics experiment one is
more inclined to treat the laser field as an internal reference frame
that must be quantized too, whereas the earth or the fixed stars are 
typically
treated as a classical, external, reference frame.
However, as a matter of principle, there is {\em no} difference
between those two cases, and in both cases one has a choice
whether to internalize or externalize the reference frame.

In short, if one has the point of view that a singlet state 
of two spin-entangled electrons or two polarization-entangled photons
can display ``2-particle nonlocality,''
then it is just as valid to claim 
that the state (\ref{one}) can display ``single-particle nonlcality.''
$|0\rangle|1\rangle+|1\rangle|0\rangle$ is entangled.

I thank Bj\"orn Hessmo,
Myungshik Kim and Howard Wiseman
for their useful comments, and Lev Vaidman for bringing
to my attention his work on this issue.


\begin{thebibliography}{99}
\bibitem{electrons}Actually, the argument would work just as well
in principle with electrons instead
of photons, as electrons, too, can be used to excite atoms.
\end{thebibliography}
\end{document}